\documentclass{ws-mpla}
\usepackage[super]{cite}
\usepackage{graphicx}
\usepackage{float}
\newcommand\rf[1]{(\ref{eq:#1})}
\newcommand\lab[1]{\label{eq:#1}}
\newcommand\nonu{\nonumber}
\newcommand\breq{\begin{eqnarray}}
\newcommand\er{\end{eqnarray}}
\newcommand\be{\begin{equation}}
\newcommand\ee{\end{equation}}

\newcommand\lb{\lbrack}
\newcommand\rb{\rbrack}

\renewcommand\({\left(}
\renewcommand\){\right)}

\newcommand\bc{\begin{center}}
\newcommand\ec{\end{center}}

\renewcommand\d{\delta}

\newcommand\vareps{\varepsilon}

\newcommand\h{\frac{1}{2}}
\renewcommand\k{\kappa}
\renewcommand\l{\lambda}
\renewcommand\L{\Lambda}
\newcommand\m{\mu}
\newcommand\n{\nu}
\newcommand\om{\omega}

\renewcommand\P{\Phi}
\newcommand\pa{\partial}

\newcommand\pr{\prime}

\newcommand\s{\sigma}

\newcommand\wti{\widetilde}

\newcommand\cC{{\mathcal C}}

\newcommand\cL{{\mathcal L}}
\newcommand\cM{{\mathcal M}}

\newcommand{\ct}[1]{\cite{#1}}

\newcommand\PRD[3]{\textsl{Phys. Rev.} \textbf{D#1}, #3 (#2)}

\newcommand\PLB[3]{\textsl{Phys. Lett.} \textbf{#1B}, #3 (#2)}
\newcommand\CQG[3]{\textsl{Class. Quantum Grav.} \textbf{#1}, #3 (#2)}
\newcommand\PRep[3]{\textsl{Phys. Reports} \textbf{#1}, #3 (#2)}

\newcommand\IJMPA[3]{\textsl{Int. J. Mod. Phys.} \textbf{A#1}, #3 (#2)}
\newcommand\IJMPD[3]{\textsl{Int. J. Mod. Phys.} \textbf{D#1}, #3 (#2)}

\newcommand\JPA[3]{\textsl{J. Physics} \textbf{A#1}, #3 (#2)}

\newcommand\phidot{\stackrel{.}{\phi}}
\newcommand\phiddot{\stackrel{..}{\phi}}
\newcommand\chidot{\stackrel{.}{\chi}}
\newcommand\chiddot{\stackrel{..}{\chi}}

\newcommand\rhodot{\stackrel{.}{\rho}}

\newcommand\adot{\stackrel{.}{a}}
\newcommand\addot{\stackrel{..}{a}}
\newcommand\Cdot{\stackrel{.}{C}}
\newcommand\Hdot{\stackrel{.}{H}}

\begin{document}

\markboth{E. Guendelman, E. Nissimov and S. Pacheva}
{Gauss-Bonnet Gravity in $D=4$ Without Gauss-Bonnet Coupling to Matter}

\catchline{}{}{}{}{}

\title{GAUSS-BONNET GRAVITY IN $D=4$ WITHOUT GAUSS-BONNET COUPLING TO MATTER 
-- COSMOLOGICAL IMPLICATIONS}

\author{\footnotesize Eduardo Guendelman}

\address{Department of Physics, Ben-Gurion University of the Negev \\
P.O.Box 653, IL-84105 ~Beer-Sheva, Israel \\
guendel@bgu.ac.il}

\author{\footnotesize Emil Nissimov and Svetlana Pacheva}

\address{Institute for Nuclear Research and Nuclear Energy, Bulgarian Academy
of Sciences\\
Boul. Tsarigradsko Chausee 72, BG-1784 ~Sofia, Bulgaria \\
nissimov@inrne.bas.bg, svetlana@inrne.bas.bg}

\maketitle

\pub{Received (Day Month Year)}{Revised (Day Month Year)}

\begin{abstract}
We propose a new model of $D=4$ Gauss-Bonnet gravity. To avoid the usual property
of the integral over the standard $D=4$ Gauss-Bonnet scalar becoming a total
derivative term, we employ the formalism of metric-independent 
non-Riemannian spacetime volume elements which makes the $D=4$ Gauss-Bonnet 
action term non-trivial without the need to couple it to matter fields 
unlike the case of ordinary $D=4$ Gauss-Bonnet gravity models. 
The non-Riemannian volume element dynamically triggers {\em the Gauss-Bonnet
scalar to be an arbitrary integration constant} $M$ on-shell, which in turn has 
several interesting cosmological implications: 
(i) It yields specific solutions for the Hubble
parameter and the Friedmann scale factor as functions of time, which are
completely independent of the matter dynamics, 
{\em i.e.}, there is no back reaction by matter on the cosmological metric. 
(ii) For $M>0$ it predicts a ``coasting''-like evolution immediately after the 
Big Bang, and it yields a late universe with dynamically produced 
dark energy density given through $M$;
(iii) For the special value $M=0$ 
we obtain exactly a linear ``coasting'' cosmology; 
(iv) For $M<0$ we have in addition to the Big Bang also a Big Crunch with 
``coasting''-like evolution around both;
(v) It allows for an explicit analytic 
solution of the pertinent Friedmann and $\phi$ scalar field equations of 
motion, while dynamically fixing uniquely the functional dependence on 
$\phi$ of the scalar potential.

\end{abstract}

\ccode{PACS Nos.: 04.50.Kd, 
98.80.Jk, 
95.36.+x.} 

\section{Introduction}

Extended gravity theories as alternatives/generalizations of the standard
Einstein General Relativity (for detailed accounts, see 
Ref.~\refcite{extended-grav,extended-grav-book,odintsov-1,odintsov-2} and
references therein) enjoy a very active development
in the last decade or so due to pressing motivation from various areas:
cosmology (problems of dark energy and dark matter), quantum field theory 
in curved spacetime (renormalization in higher loops), string theory
(low-energy effective field theories).

When considering alternative/extended theories to General Relativity, one
option is to employ alternative non-Riemannian spacetime volume-forms 
(metric-independent generally covariant volume elements or integration measure 
densities) in the pertinent Lagrangian actions instead of the canonical 
Riemannian one given by the square-root of the determinant of the Riemannian metric.

To this end let us briefly recall the  essential features of the formalism 
of non-Riemannian spacetime volume-forms, which are defined in
terms of auxiliary antisymmetric tensor gauge fields of maximal rank.
This formalism is the basis for constructing a series of extended gravity-matter
models describing unified dark energy and dark matter scenario \ct{EJP},
quintessential cosmological models with gravity-assisted and inflaton-assisted
dynamical generation or suppression of electroweak spontaneous symmetry
breaking and charge confinement \ct{grf-essay,varna-17,bpu-10}, and a novel 
mechanism for the supersymmetric Brout-Englert-Higgs effect in supergravity 
\ct{susyssb-1} 
(see Ref.~\refcite{susyssb-1,grav-bags} for a consistent geometrical formulation
of the non-Riemannian volume-form approach, which is an extension of the 
originally proposed method \ct{TMT-orig-1,TMT-orig-2}). 

Volume-forms (generally-covariant integration measures) in integrals over
manifolds are given by nonsingular maximal rank differential forms $\om$:
\breq
\int_{\cM} \om \bigl(\ldots\bigr) = \int_{\cM} dx^D\, \Omega \bigl(\ldots\bigr)
\;\; ,\;\; 
\om = \frac{1}{D!}\om_{\m_1 \ldots \m_D} dx^{\m_1}\wedge \ldots \wedge dx^{\m_D}\; ,
\lab{omega-1} \\
\om_{\m_1 \ldots \m_D} = - \vareps_{\m_1 \ldots \m_D} \Omega \;\; ,\;\;
dx^{\m_1}\wedge \ldots \wedge dx^{\m_D} = \vareps^{\m_1 \ldots \m_D}\,  dx^D \; ,
\lab{omega-3}
\er
(our conventions for the alternating symbols $\vareps^{\m_1,\ldots,\m_D}$ and
$\vareps_{\m_1,\ldots,\m_D}$ are: $\vareps^{01\ldots D-1}=1$ and
$\vareps_{01\ldots D-1}=-1$).
The volume element (integration measure density) $\Omega$ transforms as scalar
density under general coordinate reparametrizations.

In standard generally-covariant theories (with action 
$S=\int d^D\! x \sqrt{-g} \cL$)
the Riemannian spacetime volume-form is defined through the ``D-bein''
(frame-bundle) canonical one-forms $e^A = e^A_\m dx^\m$ ($A=0,\ldots ,D-1$),
related to the Riemannian metric ($g_{\m\n} = e^A_\m e^B_\n \eta_{AB}$ , 
$\eta_{AB} \equiv {\rm diag}(-1,1,\ldots,1)$):
\breq
\om = e^0 \wedge \ldots \wedge e^{D-1} = \det\Vert e^A_\m \Vert\,
dx^{\m_1}\wedge \ldots \wedge dx^{\m_D} 
\nonu \\
\longrightarrow \quad
\Omega = \det\Vert e^A_\m \Vert\, d^D x = \sqrt{-\det\Vert g_{\m\n}\Vert}\, d^D x
\; .
\lab{omega-riemannian}
\er

We will employ (Section 2 below) instead of $\sqrt{-g}$ another
alternative {\em non-Riemannian} volume element as in \rf{omega-1}-\rf{omega-3}
given by a non-singular {\em exact} $D$-form $\om = d \cC$ where:
\be
\cC = \frac{1}{(D-1)!} C_{\m_1\ldots\m_{D-1}} 
dx^{\m_1}\wedge\ldots\wedge dx^{\m_{D-1}} \; ,
\lab{C-form}
\ee
so that the {\em non-Riemannian} volume element reads:
\be
\Omega \equiv \Phi(C) = 
\frac{1}{(D-1)!}\vareps^{\m_1\ldots\m_D}\, \pa_{\m_1} C_{\m_2\ldots\m_D} \; .
\lab{Phi-D}
\ee
Here $C_{\m_1\ldots\m_{D-1}}$ is an auxiliary rank $(D-1)$ antisymmetric tensor 
gauge field. $\Phi(C)$, which is in fact the density of the dual of the rank 
$D$ field-strength
$F_{\m_1 \ldots \m_D} = \frac{1}{(D-1)!} \pa_{\lb\m_1} C_{\m_2\ldots\m_D\rb}
= - \vareps_{\m_1 \ldots \m_D} \P (C)$, 
similarly transforms as scalar density under general coordinate reparametrizations.
Let us note that the non-Riemannian volume element $\P (C)$ is dimensionless
like the standard Riemannian one $\sqrt{-g}$, meaning that the auxialiary 
antisymmetric tensor gauge field $C_{\m_1\ldots\m_{D-1}}$ has dimension 1 in
units of length.

Now, it is clear that if we replace the usual Riemannian volume element
$\sqrt{-g}$ with a non-Riemannian one $\P(C)$ in the Lagrangian action integral 
over the Gauss-Bonnet scalar $\int d^4 x\,\P(C)\,R^2_{\rm GB}$ 
(Eqs.\rf{TMT-GB}-\rf{GB-def} below), then the latter
will cease to be a total derivative in $D=4$. Thus, we will avoid the
necessity to couple $R^2_{\rm GB}$ in $D=4$ directly to matter fields 
or to use nonlinear functions of $R^2_{\rm GB}$
unlike the usual $D=4$ Gauss-Bonnet gravity 
-- for reviews, see Ref.~\refcite{GB-grav-3,GB-grav-4}; for recent discussions of
Gauss-Bonnet cosmology, see
Ref.~\refcite{GB-grav-cosmolog-1}-\refcite{GB-grav-cosmolog-10})
and references therein.

The main new feature, displayed in Section 2, of our non-standard $D=4$ 
Gauss-Bonnet gravity with a Gauss-Bonnet action term 
$\int d^4 x\,\P(C) R^2_{\rm GB}$ is due to the
equation of motion w.r.t. auxiliary tensor gauge field defining $\P(C)$
as in \rf{Phi-D}, namely it dynamically triggers the Gauss-Bonnet scalar
$R^2_{\rm GB}$ to be on-shell an arbitrary integration constant
(Eq.\rf{GB-const} below).
The latter property has, however, a consequence -- now the composite field 
$\chi = \frac{\P(C)}{\sqrt{-g}}$ appears as an additional physical field
degree of freedom related to the geometry of spacetime and its role in the
cosmological setting is described below. Let us note that this is in sharp
contrast w.r.t. other extended gravity-matter models constructed in terms of 
(one or several) non-Riemannian volume forms 
\ct{susyssb-1,grav-bags,EJP,grf-essay,varna-17,bpu-10}, where we start within 
the first-order formalism and where
composite fields of the type of $\chi$ 
turn out to be (almost) pure gauge (non-propagating) degrees of freedom.

The dynamically triggered constancy of $R^2_{\rm GB}$ in turn has
several interesting implications for cosmology. 

As we will show in Section 3 below,
the cosmological dynamics in the new $D=4$ Gauss-Bonnet gravity provides
automatically a ``coasting'' evolution of the early universe near the Big
Bang at $t=t_{\rm BB}$, where the Hubble parameter $H(t) \sim (t-t_{\rm BB})^{-1}$ 
and the Friedmann scale factor $a(t) \sim (t-t_{\rm BB})$, \textsl{i.e.}, 
space size $a(t)$ and horizon size $H^{-1}(t)$ expand at the same rate 
(no horizon problem); for a general discussion of ``coasting'' cosmological 
evolution, see Ref.~\refcite{coasting-1}-\refcite{coasting-4}.
Furthermore, for late times we obtain either a de Sitter universe with dynamically
generated dark energy density or a Big Crunch depending on the sign of the 
dynamically generated constant value of $R^2_{\rm GB}$.

An important observation here is that the cosmological solution for $H(t)$
and $a(t)$ does {\em not} feel the details of the matter content and the matter 
dynamics, \textsl{i.e.}, there is no direct back reaction of matter on the
cosmological metric. The reason here is that the differential equation determining the
solution $H(t)$ (or $a(t)$) results from the equation for the Gauss-Bonnet
scalar equalling arbitrary integration constant on-shell, which does not involve any
matter terms. On the other hand, matter terms are present in the pertinent
Friedmann equations (Section 3) which now reduce to a differential equation for
the composite field $\chi = \P(C)/\sqrt{-g}$. Therefore, the solution for 
$\chi$ (Eq.\rf{chi-sol} below) completely absorbs the impact of the matter dynamics while the 
overall solution for $H(t)$ and $a(t)$ is left unchanged. Furthermore, if we
``freeze'' $\chi$ to be a constant (Section 4 below), then in the case of scalar 
field $\phi$ matter the exact expression for the corresponding scalar potential 
$V(\phi)$ as function of $\phi$ is fixed uniquely.

The above described main properties of the present version of $D=4$
Gauss-Bonnet gravity, namely, the dynamical constancy of the Gauss-Bonnet
scalar derived from a Lagrangian action principle and the appearance of an
additional degree of freedom $\chi$ absorbing the effect of the matter dynamics,
are the most significant differences w.r.t. the approach in several recent
papers \ct{myrzakulov,vagenas,radinschi} extensively studying static spherically 
symmetric solutions in
gravitational theories in the presence of a constant Gauss-Bonnet scalar, 
where the constancy of the latter is {\em imposed} as an additional condition 
on-shell beyond the standard equations of motion resulting from an action principle.

In the last discussion section we point out a limitation of the present non-canonical
$D=4$ Einstein-Gauss-Bonnet model, namely, that it predicts continuous acceleration 
throughout the whole evolution of the universe, and we briefly describe a generalization 
of our model allowing for both acceleration and deceleration.

\section{Gauss-Bonnet Gravity in $D=4$ With a Non-Riemannian Volume Element}

We propose the following self-consistent action of $D=4$ Gauss-Bonnet
gravity without the need to couple the Gauss-Bonnet scalar to some matter
fields
(for simplicity we are using units with the Newton constant $G_N = 1/16\pi$):
\be
S = \int d^4 x \sqrt{-g} \Bigl\lb R + L_{\rm matter}\Bigr\rb
+ \int d^4x \,\P (C)\, R_{\rm GB}^2 \; .
\lab{TMT-GB}
\ee
Here the notations used are as follows (we employ the usual second order
formalism):
\begin{itemize}
\item
$R_{\rm GB}^2$ denotes the Gauss-Bonnet scalar:
\be
R_{\rm GB}^2 \equiv R^2 - 4 R_{\m\n} R^{\m\n} + R_{\m\n\k\l} R^{\m\n\k\l} \; .
\lab{GB-def}
\ee
\item
$\P (C)$ denotes a non-Riemannian volume element defined as a scalar density
of the dual field-strength of an auxiliary antisymmetric tensor gauge field of 
maximal rank $C_{\m\n\l}$:
\be
\P (C) = \frac{1}{3!} \vareps^{\m\n\k\l} \pa_\m C_{\n\k\l} \; .
\lab{PC-def}
\ee
Let us particularly stress that, although we stay in $D=4$ spacetime
dimensions and although we {\em don't couple} the Gauss-Bonnet scalar 
\rf{GB-def} to the matter fields, the last term in \rf{TMT-GB}
thanks to the presence of the non-Riemannian volume element \rf{PC-def} is 
non-trivial ({\em not} a total derivative as with the ordinary Riemannian
volume element $\sqrt{-g}$)) and yields a non-rivial contribution to the
Einstein equations (Eqs.\rf{einstein-eqs} below).
\item
As a matter Lagrangian we will take for simplicity an ordinary scalar field
one:
\be 
L_{\rm matter} = - \h g^{\m\n} \pa_\m \phi \pa_\n \phi - V(\phi) \; .
\lab{matter-1}
\ee 
As discussed below, the explicit choice of $L_{\rm matter}$ does not affect
the cosmological solutions for the Hubble parameter and the Friedmann scale
factor. It only affects the solution for the composite field $\chi$ defined
in Eq.\rf{chi-def} (cf. Eq.\rf{chi-sol} below).

\end{itemize}

We now have three types of equations of motion resulting from the action
\rf{TMT-GB}: 
\begin{itemize}
\item
Einstein equations w.r.t. $g^{\m\n}$ where we employ the definition for a
dimensionelss
composite field:
\be
\chi \equiv \frac{\P(C)}{\sqrt{-g}} 
\lab{chi-def}
\ee
representing the ratio of the non-Riemannian to the standard Riemannian 
volume element:
\breq
R_{\m\n} - \h g_{\m\n}R = \h T_{\m\n} - \h g_{\m\n} \chi R_{\rm GB}^2
+ 2R \nabla_\m \nabla_\n \chi 
\nonu \\
+ 4 \Box \chi \bigl( R_{\m\n} - \h g_{\m\n}R \bigr)
- 4 R_{\m}^{\rho} \nabla_\rho \nabla_\n \chi
- 4 R_{\n}^{\rho} \nabla_\rho \nabla_\m \chi 
\nonu \\
+ 4 g_{\m\n} R^{\rho \s} \nabla_\rho \nabla_\s \chi 
- 4 g^{\k\rho} g^{\l\s} R_{\m\k\n\l} \nabla_\rho \nabla_\s \chi \; , 
\lab{einstein-eqs}
\er
where 
$T_{\m\n} = g_{\m\n} L_{\rm matter} - 2\frac{\pa}{\pa g^{\m\n}}L_{\rm matter}$ 
is the standard matter energy-momentum tensor:
\be
T_{\m\n} = \pa_\m \phi \pa_\n \phi 
- g_{\m\n} \Bigl( \h g^{\k\l} \pa_\k \phi \pa_\l \phi + V(\phi)\Bigr) \; .
\lab{T-def}
\ee
\item
The equations of motion w.r.t. scalar field $\phi$ have the standard form 
(they are not affected by the presence of the Gauss-Bonnet term):
\be
\Box \phi + \frac{\pa V}{\pa \phi} = 0 \; .
\lab{vp-eq} 
\ee
\item
The crucial new feature are the equations of motion w.r.t. auxiliary
non-Riemannian volume element tensor gauge field $C_{\m\n\l}$:
\be
0 = \frac{\d}{\d C_{\n\k\l}} \int d^4 x \P(C) R^2_{\rm GB} = 
- \frac{1}{3!} \vareps^{\m\n\k\l} \pa_\m R^2_{\rm GB}  \; ,
\lab{C-eq}
\ee
that is:
\be
\pa_\m R_{\rm GB}^2 = 0 \quad \longrightarrow \quad
R_{\rm GB}^2 = 24 M = {\rm const} \; ,
\lab{GB-const}
\ee
where $M$ is an arbitrary dimensionful integration constant and 
the numerical factor 24 in \rf{GB-const} is chosen for later convenience.
\end{itemize}

The dynamically triggered constancy of the Gauss-Bonnet scalar \rf{GB-const}
comes at a price as we see from the generalized Einstein
Eqs.\rf{einstein-eqs} -- namely, now the composite field 
$\chi = \frac{\P(C)}{\sqrt{-g}}$ appears as an additional physical field
degree of freedom.

In what follows we will see (Section 4) that when considering cosmological 
solutions we can consistently ``freeze'' the
composite field $\chi = {\rm const}$ so that all terms on
the r.h.s. of \rf{einstein-eqs} with derivatives of the composite field $\chi$
will vanish. The freezing of $\chi$ together with \rf{GB-const} has two main 
effects:

(a) It produces on r.h.s. of \rf{einstein-eqs} a dynamically generated 
cosmological constant $\L_0$-term:
\be
- g_{\m\n} \chi R_{\rm GB}^2 = - g_{\m\n} 2 \L_0 \;\;,\;\;
\L_0 \equiv 12 \chi M \; .
\lab{GB-CC}
\ee

(b) Within the class of cosmological solutions, as shown in Section 4 below,
the ``freezing'' of $\chi$
produces an explicit analytic solution of the extended Friedmann equations
and of the $\phi$ scalar field equations of motion, with simultaneous 
dynamical fixing uniquely of the functional dependence on $\phi$ of the 
scalar potential $V(\phi)$. In particular it yields the exact value of the 
vacuum energy density in the ``late'' universe -- the dark energy density 
(Eq.\rf{DE} below) -- in terms of the dynamically generated constant value 
of the Gauss-Bonnet scalar \rf{GB-const}.

\section{Cosmological Solutions with a Dynamically Constant Gauss-Bonnet Scalar}

Now we perform a Friedmann-Lemaitre-Robertson-Walker (FLRW) reduction of
the original action \rf{TMT-GB} with FLRW metric:
\be
ds^2 = g_{\m\n} dx^\m dx^\n = - N^2(t) dt^2 + a^2(t) d{\vec x}^2 
\lab{FLRW-metric}
\ee
where:
\be
\P(C) = \Cdot \quad ,\quad 
R_{\rm GB}^2 = \frac{8}{Na^3} \frac{d}{dt}\Bigl(\frac{\adot^3}{N^3}\Bigr) \; ,
\lab{GB-FLRW}
\ee
the overdot denoting $\frac{d}{dt}$. The corresponding FLRW action reads:
\be
S_{\rm FLRW} = \int dt \Bigl\{ - 6 \frac{a \adot^2}{N} 
+ Na^3 \bigl\lb \frac{\phidot^2}{2N^2} - V(\phi)\bigr\rb
+ \frac{8 \Cdot}{Na^3} \frac{d}{dt}\Bigl(\frac{\adot^3}{N^3}\Bigr)\Bigr\} \; .
\lab{TMT-GB-FLRW}
\ee
Accordingly, the equations of motion of \rf{TMT-GB-FLRW} -- FLRW counterparts 
of Eqs.\rf{GB-const},\rf{einstein-eqs},\rf{vp-eq} -- acquire the form (as usual,
after variation w.r.t. lapse function $N(t)$ we set the gauge $N(t)=1$):
\begin{itemize}
\item
The FLRW counterpart of \rf{GB-const} (the constancy of Gauss-Bonnet scalar) 
becomes:
\be
\frac{\adot^2}{a^2}\, \frac{\addot}{a} = M \quad \longrightarrow \quad
\Hdot = - H^2 + \frac{M}{H^2} \; ,
\lab{GB-FRLW-const}
\ee
$H = \frac{\adot}{a}$ denoting the Hubble parameter. It is important to stress that 
the FLRW spacetime geometry, {\em i.e.}, $a=a(t)$ is completely determined by the
solution of Eq.\rf{GB-FRLW-const} (see Eqs.\rf{H-integr}-\rf{H-sol} below)
and it {\em does not feel any back reaction} by the matter fields. 
\item
The FLRW counterparts of the generalized Einstein Eqs.\rf{einstein-eqs}, 
{\em i.e.}, the Friedmann equations become upon using Eq.\rf{GB-FRLW-const}:
\breq
6H^2 - (\rho + 24 M \chi) + 24\chidot H^3 = 0 \; ,
\lab{Fried-1} \\
6H^2 + 3(p - 24 M \chi) + \frac{12 M}{H^2} 
+ 24\bigl(\chiddot H^2 - 2 \chidot \frac{M}{H}\bigr) = 0 \; ,
\lab{Fried-2}
\er
where now $\chi \equiv \frac{\Cdot}{a^3}$ and the energy density $\rho$ and
pressure $p$ have the usual form:
\be
\rho = \h \phidot^2 + V(\phi) \quad ,\quad p = \h \phidot^2 - V(\phi) \; .
\lab{rho-p-def}
\ee
\item
The FLRW $\phi$-equation and the corresponding energy-conservation equation
have the ordinary form:
\be
\phiddot + 3 H \phidot + \frac{\pa V}{\pa \phi} = 0 \quad ,\quad
3H (\rho + p) + \rhodot = 0 \; .
\lab{vp-rho-p-eqs}
\ee
Thus, while $H(t)$ (and $a(t)$) do not feel back reaction from the matter fields, 
they in turn significantly impact the matter fields' dynamics. 
\end{itemize}

Compatibility between the two Friedmann Eqs.\rf{Fried-1}-\rf{Fried-2} can be
explicitly checked to follow from the second Eq.\rf{vp-rho-p-eqs}.

The first Friedmann Eq.\rf{Fried-1} can be represented as a differential equation
for $\chi(t)$:
\be
\chidot = \frac{M}{H^3}\,\chi + \frac{\rho}{24 H^3} - \frac{1}{4H} \; ,
\lab{chi-eq}
\ee
whose solution is given through the solutions for $H(t)$ from \rf{GB-FRLW-const}
and matter energy density $\rho$ \rf{rho-p-def} from \rf{vp-rho-p-eqs}:
\be
\chi(t) = 
e^{\int dt^\pr \frac{M}{H^3}} \Bigl\{\int dt^{\pr\pr} \Bigl\lb 
\Bigl(\frac{\rho}{24 H^3}-\frac{1}{4H}\Bigr) e^{-\int dt^\pr \frac{M}{H^3}}\Bigr\rb
+ {\rm const}\Bigr\} \; .
\lab{chi-sol}
\ee
Eq.\rf{chi-sol} shows that it is $\chi$ which absorbs the backreaction of
matter unlike the Friedmann scale factor $a(t)$ or Hubble parameter $H(t)$, 
since Eq.\rf{GB-FRLW-const} does not involve matter.
In fact we could take in the Friedmann equations \rf{Fried-1}, 
{\em i.e.}, \rf{chi-eq}, and \rf{Fried-2} any kind of matter which obeys
the covariant energy conservation (second Eq.\rf{vp-rho-p-eqs}).


Now, we observe that due to second Eq.\rf{GB-FRLW-const} -- the dynamical 
constancy of the Gauss-Bonnet scalar -- there is permanent
acceleration/deceleration throughout the whole evolution of the universe:
\be
\frac{\addot}{a} = \Hdot + H^2 = \frac{M}{H^2}  
\lab{accel}
\ee
depending on the sign of $M$. We will first assume the integration constant
$M>0$, \textsl{i.e.}, permanent acceleration. The other cases 
($M=0$ -- permanent ``coasting'', and $M<0$ -- permanent deceleration)
will be briefly discussed in the next subsections below.

\subsection{Friedmann Scale Factor Solution for $M>0$}

We can solve explicitly the second Eq.\rf{GB-FRLW-const} -- simple differential 
equation for $H=H(t)$:
\be
\int \frac{dH}{\frac{M}{H^2} - H^2} = t-t_0 \; ,
\lab{H-integr}
\ee
where $t_0$ is an integration constant, \rf{H-integr} yielding:
\be
4 M^{1/4} \bigl(t - t_{\rm BB}\bigr) = 
\log\Bigl(\frac{H(t)+M^{1/4}}{H(t)-M^{1/4}}\Bigr)
+\pi - 2\arctan\Bigl(\frac{H(t)}{M^{1/4}}\Bigr) \; ,
\lab{H-sol}
\ee
with:
\be
t_{\rm BB} \equiv t_0 - \frac{\pi}{4 M^{1/4}} \; .
\lab{t-BB}
\ee
The solution $H(t)$ implicitly defined in \rf{H-sol} is graphically depicted
in Fig.1.

\begin{figure}[H]
\begin{center}
\includegraphics[width=7cm,keepaspectratio=true]{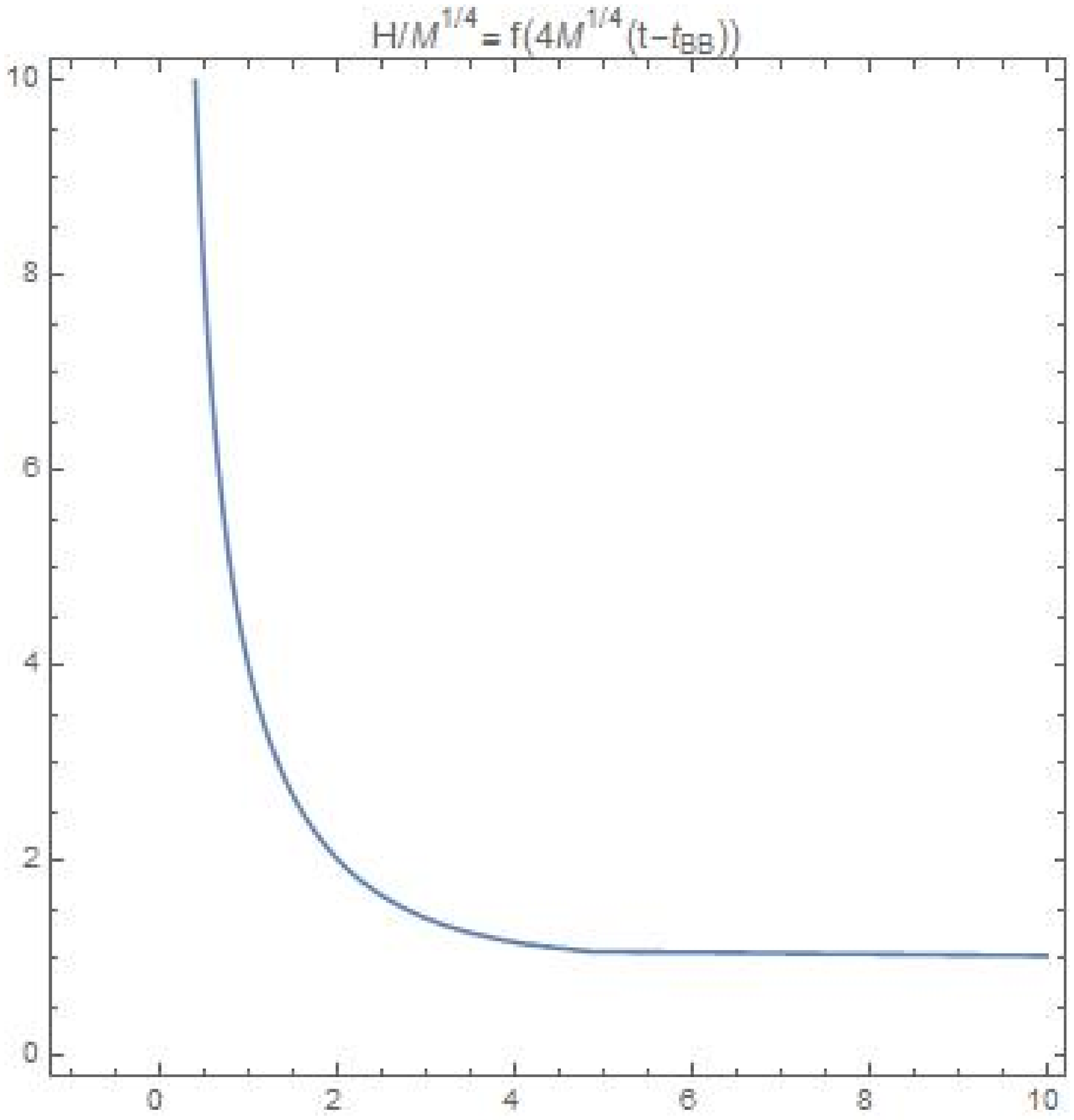}
\caption{$H(t)$, implicitly defined in Eq.\rf{H-sol}, 
as function of $(t-t_{\rm BB})$.}
\end{center}
\end{figure}

First, we notice from Eq.\rf{H-sol} that:
\be
H(t) \to \infty \quad {\rm for}\;\; t-t_{\rm BB} \to 0 \; ,
\lab{H-to-BB}
\ee
in other words, there is a {\em Big Bang} at $t = t_{\rm BB}$ \rf{t-BB}.
Expanding the r.h.s. of Eq.\rf{H-sol} for small $(t-t_{\rm BB})$, equivalent
to expanding for small $M$, yields:
\breq
t-t_{\rm BB} = \frac{1}{H} + \frac{M}{5 H^5} + {\rm O}(M^2) 
\;\; \longrightarrow \;\;
\nonu \\
H(t) = \frac{1}{t-t_{\rm BB}} + \frac{M}{5} (t-t_{\rm BB})^3 + {\rm O}(M^2)
\; .
\lab{H-coasting}
\er
The last relation in\rf{H-coasting} implies a ``coasting'' \ct{coasting-1}
behavior of the universe for evolution times near the Big Bang. 
For the Friedmann scale factor itself we get from \rf{H-coasting}:
\breq
a(t)/a_0 = (t-t_{\rm BB}) \exp\bigl\{\frac{M}{20}(t-t_{\rm BB})^4 + 
{\rm O}(M^2)\bigr\} \; ,
\lab{a-coasting} \\
{\rm i.e.} \;\; a(t)/a_0 \simeq t-t_{\rm BB} \;\; {\rm for}\;
{\rm small}\; (t-t_{\rm BB}) \; ,
\nonu
\er
in other words, close to the Big Bang the horizon $H^{-1}(t)$ and space $a(t)$ 
evolve in the same way.

On the other hand, for large $t$ (``late'' universe):
\be
H(t) \simeq M^{1/4} \quad ,\quad \frac{\addot}{a}\simeq \sqrt{M} = 
\frac{1}{3} \L_{\rm DE} \; ,
\lab{H-late}
\ee
\textsl{i.e.}, the universe evolves with a constant acceleration where
$2 \L_{\rm DE} \equiv 6 \sqrt{M}$ is the dark energy density.

\subsection{Friedmann Scale Factor Solution for $M=0$}

Let us consider the special case $M=0$, {\em i.e.},
according to \rf{GB-const} $R_{\rm GB}^2$ dynamically vanishes.
From Eq.\rf{GB-FRLW-const} 
we have:
\be
\Hdot = - H^2 \;\; \to \;\; H(t) = \frac{1}{t-t_0} \;\; \to \;\;
a(t)/a_0 = t-t_0 \;,
\lab{GB-FRLW-const-0}
\ee
which represent {\em ``coasting''} universe' evolution (linear expansion of the
Friedmann scale factor) for all the time after the Big Bang at $t=t_0$.

Let us note that dynamical vanishing of $D=4$ Gauss-Bonnet scalar has been
previously obtained in Ref.~\refcite{zanelli}, however within a different and
physically non-equivalent formalism - first-order (frame-bundle) formalism with 
a nonvanishing torsion.

\subsection{Friedmann Scale Factor Solution for $M<0$}

Now we consider the case $M<0$ (we will use the symbol ${\bar M} \equiv - M$
for convenience, {\em i.e.}, according to \rf{GB-const} 
$R_{\rm GB}^2 = - 24 {\bar M} <0$). Setting in the integral \rf{H-integr}
$M = - {\bar M}$ we obtain the implicit solution for $H={\bar H}(t)$:
\breq
4\sqrt{2}{\bar M}^{1/4} (t-t_0) =
\log\Bigl(\frac{{\bar H}^2(t) + {\bar M}^{1/2} + 
\sqrt{2}{\bar M}^{1/4} {\bar H}(t)}{{\bar H}^2(t) + {\bar M}^{1/2} 
- \sqrt{2}{\bar M}^{1/4} {\bar H}(t)}\Bigr)
\nonu \\
- 2 \arctan\Bigl(\frac{\sqrt{2}{\bar H}(t)}{{\bar M}^{1/4}}-1\Bigr) 
- 2 \arctan\Bigl(\frac{\sqrt{2}{\bar H}(t)}{{\bar M}^{1/4}}+1\Bigr)\;,
\;\;\;
\lab{H-sol-bar}
\er
where the integration constant $t_0$ is defined by ${\bar H}(t=t_0) = 0$. 
From \rf{H-sol-bar} we find both a Big Bang:
\be
{\bar H}(t) \to +\infty \;\; {\rm for} \; 
t= t_{\rm BB}\equiv t_0 - \frac{\pi}{2\sqrt{2}{\bar M}^{1/4}} \; ,
\lab{big-bang-bar} 
\ee
and a Big Crunch at finite cosmological times:
\be
{\bar H}(t) \to -\infty \;\; {\rm for} \; 
t= t_{\rm BC}\equiv t_0 + \frac{\pi}{2\sqrt{2}{\bar M}^{1/4}} \;.
\lab{big-crunch-bar}
\ee
Similarly to the case $M>0$ \rf{H-coasting}-\rf{a-coasting} we obtain a 
``coasting'' behaviour near the Big Bang ($(t-t_{\rm BB})$ small):
\breq
{\bar H}(t) = \frac{1}{(t-t_{\rm BB})} - \frac{{\bar M}}{5} (t-t_{\rm BB})^3
+ {\rm O}({\bar M}^2) \; ,
\lab{H-bar-coasting} \\
a(t)/a_0 = (t-t_{\rm BB}) \exp\bigl\{ -\frac{{\bar M}}{20} (t-t_{\rm BB})^4
{\rm O}({\bar M}^2)\bigr\} \; .
\lab{a-bar-coasting}
\er
Near the Big Crunch ($(t_{\rm BC}-t)$ small) there is also a ``coasting''
behavior:
\breq
{\bar H}(t) = -\frac{1}{(t_{\rm BC}-t)} + \frac{{\bar M}}{5} (t_{\rm BC}-t)^3
+ {\rm O}({\bar M}^2) \; ,
\lab{H-bar-crunch} \\
a(t)/a_0 = (t_{\rm BC}-t) \exp\bigl\{ -\frac{{\bar M}}{20} (t_{\rm BC}-t)^4 
+ {\rm O}({\bar M}^2)\bigr\} \; .
\lab{a-bar-crunch}
\er
Here once again we observe that both near the Big Bang and near the Big
Crunch the horizon and space sizes evolve in the same way.

\section{Special Cosmological Solution of the full
Gauss-Bonnet Gravity with a Dynamically Constant Gauss-Bonnet Scalar}

We will now study in some detail special particular solutions of the
extended Einstein Eqs.\rf{einstein-eqs} (taking into account \rf{GB-const}) 
with a {\em frozen} composite field $\chi$ \rf{chi-def} ($\chi = {\rm const}$):
\be
R_{\m\n} - \h g_{\m\n}R = \h T_{\m\n} - 12 g_{\m\n} \chi M \quad ,\quad
R^2_{\rm GB} = 24 M \; .
\lab{einstein-eqs-chi-const}
\ee
Let us stress that we take the composite field 
$\chi\equiv \frac{\P(C)}{\sqrt{-g}}$ to be ``frozen'' to a constant
as an {\em additional} condition {\em after} we derive the full system of 
equations of motion, including \rf{einstein-eqs} and \rf{C-eq}, from
the original non-Riemannian volume-form action \rf{TMT-GB}.

As we will see below, consistency of the system \rf{einstein-eqs-chi-const}
will require the pertinent matter Lagrangian to be of a very specific form
depending on the sign of the integration constant $M$. 

We start by looking for cosmological solutions with $\chi = {\rm const}$
of the  Eqs.\rf{Fried-1}-\rf{Fried-2} -- the FLRW reduction of 
\rf{einstein-eqs-chi-const}. 
Inserting $\chi = {\rm const}$ there we obtain:
\be
6H^2 - (\rho + 24 M \chi) = 0 \quad ,\quad
6H^2 + 3(p - 24 M \chi) + \frac{12 M}{H^2} = 0
\lab{Fried}
\ee
from which we deduce the equation of state:
\breq
{\wti \rho} + 3 {\wti p} + \frac{72M}{{\wti \rho}} = 0 \quad ,
\lab{eq-state} \\
{\wti \rho} \equiv \rho + 24 M\chi \quad ,\quad {\wti p} \equiv p - 24M\chi \; ,
\lab{shifted}
\er
where ${\wti \rho}$ and ${\wti p}$ are the shifted energy density and pressure
incorporating the dynamically Gauss-Bonnet induced cosmological constant 
$\L_0 = 12\chi M$ \rf{GB-CC}.


From the Friedmann Eqs.\rf{Fried-1}-\rf{Fried-2} we find:
\breq
\phidot^2 = \rho + p = 4 \Bigl(H^2 - \frac{M}{H^2}\Bigr) \; ,
\lab{vp-square} \\
V(\phi) = \h (\rho - p) = 4H^2 + 2\frac{M}{H^2} - 24M\chi \; ,
\lab{V-eq} \\
{\wti V}(\phi) \equiv V(\phi) + 24M\chi = 4H^2 + 2\frac{M}{H^2} \; ,
\lab{V-shifted}
\er
where ${\wti V}(\phi)$ is the shifted scalar potential incorporating the
dynamically Gauss-Bonnet induced cosmological constant $12M\chi$ \rf{GB-CC}.

\subsection{Case $M>0$}

Combining $\phi$-equations of motion \rf{vp-rho-p-eqs} and the equation of state
\rf{eq-state} yields a differential equation for ${\wti\rho}$ as function of
the Friedmann scale factor $a$:
\be
\frac{d{\wti\rho}}{da}= -\frac{2}{a}\Bigl({\wti\rho} - \frac{36M}{{\wti\rho}}\Bigr)
\quad \longrightarrow \quad {\wti\rho}(a) = \sqrt{36M + c_0 a^{-4}} \; ,
\lab{rho-a-eq}
\ee
where $c_0$ is an integration constant. In accordance with the solution for
the Friedmann scale factor in subsection 3.1, we obtain from \rf{rho-a-eq}:
\begin{itemize}
\item
${\wti\rho}(a) \simeq \sqrt{c_0} a^{-2}$ close to the Big Bang where $a \to 0$
-- ``coasting'' behavior;
\item
${\wti\rho}(a) \simeq 6\sqrt{M}$ in the late universe where $a \to \infty$ 
exponentially -- $6\sqrt{M}$ is the dark energy density conforming to the
late universe value of $H$ \rf{H-late}.
\end{itemize}

Relation \rf{vp-square} yields solution for $\phi (t)$ as function of $t$ 
through $H(t)$ as defined implicitly in \rf{H-sol}:
\breq
\phi = 2 \int \frac{dH}{\sqrt{H^2 - \frac{M}{H^2}}} \quad \longrightarrow 
\nonu \\
\phi(t) = \log\(\sqrt{\frac{H^4(t)}{M}-1} + \frac{H^2(t)}{M^{1/2}}\)
\lab{vp-sol}
\er
or, inversely, $H (t)$ as function of $\phi (t)$:
\be
H^2 = \sqrt{M} \cosh(\phi)  \; .
\lab{H-vp}
\ee
The solution $\phi(t)$ \rf{vp-sol} is graphically depicted in Fig.2.

\begin{figure}[H]
\begin{center}
\includegraphics[width=7cm,keepaspectratio=true]{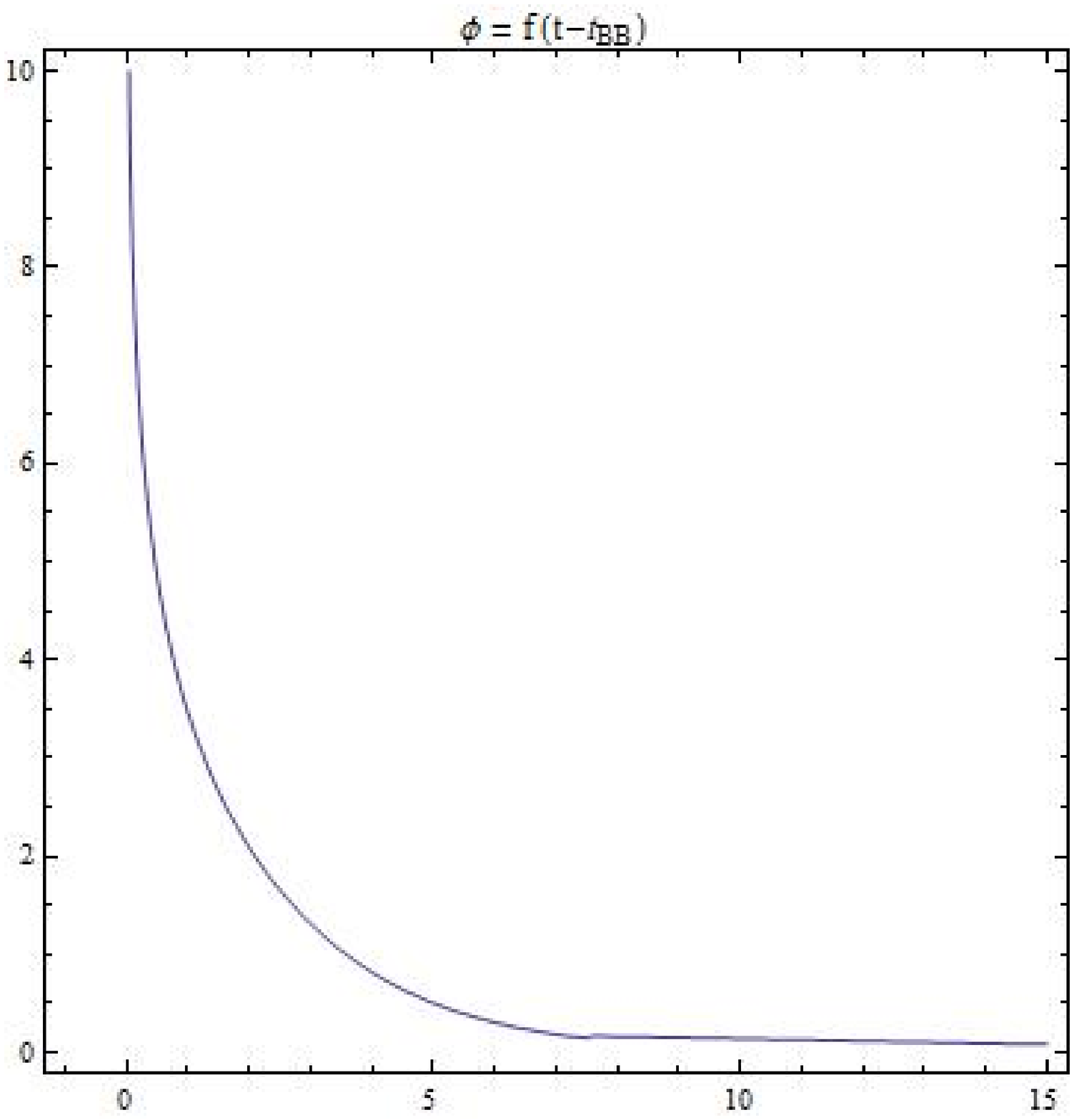}
\caption{$\phi(t)$, defined in \rf{vp-sol}, as function of $(t-t_{\rm BB})$.}
\end{center}
\end{figure}

From \rf{V-eq} and \rf{H-vp} we find that the functional dependence
of the scalar potential $V(\phi)$ or the shifted one ${\wti V}(\phi)$ 
\rf{V-shifted} is {\em uniquely fixed by the dynamics}:
\be
{\wti V}(\phi) = 4\sqrt{M} \cosh(\phi) + \frac{2\sqrt{M}}{\cosh(\phi)} \; .
\lab{V-sol}
\ee

From Eq.\rf{vp-sol} for small $(t-t_{\rm BB})$ near the Big Bang ({\em i.e.},
large $H$ according to \rf{H-coasting}) we have:
\be
\phi(t) \simeq 2\log H(t) \simeq - 2 \log (t-t_{\rm BB}) \; .
\lab{vp-coasting}
\ee
On the other hand, in the ``late'' universe $\phi(t)$ \rf{vp-sol} 
converges to zero -- the minimum of the scalar potential \rf{V-sol}, {\em i.e.}, 
which yields precisely the value of the dark energy density:
\be
{\wti V}(\phi=0)=6\sqrt{M} \; .
\lab{DE}
\ee

\subsection{Case $M=0$}

For the purely ``coasting'' cosmology (subsection 3.2)
from Eq.\rf{vp-sol} (the integral on the first line) and from Eq.\rf{V-eq} 
we obtain when $M=0$:
\be
\phi = 2\log H = - 2\log(t-t_0) \quad ,\quad V(\phi) = 4 e^{\phi} \; .
\lab{vp-V-sol-0}
\ee
The equation of state \rf{eq-state} for $M=0$ becomes:
\be
\rho + 3p = 0 \; ,
\lab{rho-3p}
\ee
which is the same as, for instance, for the string gas \ct{string-gas}.

\subsection{Case $M<0$}

Now, for the BigBang-BigCrunch cosmological solution (subsection 3.3), 
setting in the integral \rf{vp-sol} $M = - {\bar M}<0$, we obtain 
the solution for $\phi(t)$ as function of ${\bar H}(t)$ \rf{H-sol-bar}:
\be
\phi(t) = \log\(\sqrt{\frac{{\bar H}^4(t)}{{\bar M}}+1} + 
\frac{{\bar H}^2(t)}{{\bar M}^{1/2}}\)
\lab{vp-sol-bar}
\ee
or, inversely, ${\bar H}(t)$ as function of $\phi (t)$:
\be
{\bar H}^2 = \sqrt{{\bar M}} \sinh(\phi)  \; .
\lab{H-vp-bar}
\ee
Accordingly, the functional dependence of the (shifted) scalar potential for
$M=-{\bar M}<0$ is uniquely determined as:
\be
{\wti V}(\phi) = 4\sqrt{{\bar M}} \sinh(\phi) + 
\frac{2\sqrt{{\bar M}}}{\sinh(\phi)} \; .
\lab{V-sol-bar}
\ee

\section{Discussion and Outlook}

In the present paper we have made an essential use of the formalism of 
non-Riemannian spacetime volume-forms (alternative metric-independent 
volume elements, \textsl{i.e.}, generally-covariant integration measure densities) 
defined in terms of auxiliary maximal rank tensor gauge fields, 
in order to construct a new type of Einstein-Gauss-Bonnet gravity in $D=4$
avoiding the need to couple the Gauss-Bonnet scalar to any matter fields.
The presence of the non-Riemannian volume element in the $D=4$ Gauss-Bonnet
action terms makes the theory non-trivial and well-defined. The principal
new feature is that on-shell the Gauss-Bonnet scalar becomes an arbitrary
integration constant. In the cosmological setting the dynamical constancy of
the Gauss-Bonnet scalar by itself completely determines the solutions for 
the Hubble parameter $H(t)$ and the Friedmann scale factor $a(t)$
as functions of the cosmological time without any influence of the matter 
dynamics (no back reaction of matter on the cosmological metric). 

The whole effect of matter on the
spacetime is absorbed by the time-dependence of the composite field 
$\chi$ \rf{chi-def} (the ratio of the non-Riemannian volume element $\P(C)$ to the
standard Riemannian $\sqrt{-g}$). If we choose to ``freeze'' 
$\chi = {\rm const}$, then in the case of scalar field matter 
the scalar potential $V(\phi)$ is uniquely constrained to be of a very
specific form as a function of $\phi$.

The solutions for $H(t)$ and $a(t)$ between ``coasting'' early Big Bang
cosmology, where the evoluton of $a(t)$ coincides with the evolution of the
horizon ($H^{-1}(t)$), and a late time de Sitter universe ($M>0$) or a Big Crunch
($M<0$) at a finite cosmological time, depend on the sign of the dynamically
generated constant value $M$ of the Gauss-Bonnet scalar. The case $M>0$ is
more realistic, but still too simplified to be realistic enough to describe 
the whole evolution of the observed universe, since it predicts continuous  
acceleration during the whole evolution.

Thus, our $D=4$ Einstein-Gauss-Bonnet model could be considered as an 
approximation to be used in the future as a basis for a more realistic 
cosmological models. For example, a model like ours could be an interesting
candidate for the high-energy limit of a realistic model. This is because of
the property of the composite field $\chi = \P(C)/\sqrt{-g}$ to absorb the
effects of a non-trivial matter behaviour and to prevent a matter
back reaction on the spacetime metrics. Such an effect could be exactly what is 
needed to cure the issues noticed in Ref.~\refcite{afshordi}, where it has been shown 
that quantum fluctuations in the energy-momentum tensor of matter can cause serious
phenomenological problems. The latter could however be avoided provided at
high energies exists a field like $\chi$ that compensates the effects of the
matter fluctuations.

Still, we can slightly generalize our $D=4$ Einstein-Gauss-Bonnet model
\rf{TMT-GB} to yield more realistic cosmological solutions, namely, to
provide continuous acceleration during the early universe epoch after the Big
Bang as well as during the late universe (dark energy) epoch, while
providing deceleration for an intermediate epoch.

To this end we can consider the more general (than \rf{TMT-GB}) action:
\breq
S = \int d^4 x \sqrt{-g} \Bigl\lb R 
- \h g^{\m\n} \pa_\m \phi \pa_\n \phi- V(\phi) \Bigr\rb 
\nonu \\
+ \int d^4x \,\P (C)\, \Bigl(R_{\rm GB}^2 - 24 W(\phi)\Bigr) \; ,
\lab{TMT-GB-W}
\er
with an additional appropriately chosen scalar potential term $W(\phi)$ (the
factor $24$ is introduced for numerical convenience).

Now, instead of \rf{GB-const} the equations of motion w.r.t. auxiliary
tensor gauge field $C_{\m\n\l}$ defining the non-Riemannian volume element
$\P(C)$ yield:
\be
\pa_\m \Bigl(R_{\rm GB}^2 - 24 W(\phi)\Bigr) = 0 \quad \longrightarrow \quad
R_{\rm GB}^2 = 24 \bigl(M + W(\phi)\bigr) = {\rm const} \; ,
\lab{GB-const-W}
\ee
where again $M$ is an arbitrary integration constant and will be absorved
into $W(\phi)$, henceforth $W(\phi)+M \equiv {\wti W}(\phi)$.

In the cosmological FLRW reduction Eq.\rf{GB-const-W} yields instead of 
\rf{GB-FRLW-const}:
\be
\Hdot = - H^2 + \frac{{\wti W}(\phi)}{H^2} \quad ,\quad 
\frac{\addot}{a} = \frac{{\wti W}(\phi)}{H^2} \; ,
\lab{GB-FRLW-W}
\ee
so that now we can have both acceleration or deceleration.

Indeed, now upon setting as above the composite field $\chi = {\rm const}$
the Friedman equations and the equation of state, as well as the equations
for $\phidot^2$ and the initial scalar potential $V(\phi)$ retain the same form as 
in \rf{Fried}-\rf{shifted} and \rf{vp-square}-\rf{V-shifted}, respectively,
with replacing there $M \longrightarrow {\wti W}(\phi)$, in particular:
\breq
\phidot = - 2\sqrt{H^2 - \frac{{\wti W}(\phi)}{H^2}} \quad , \quad
({\rm or, ~ equivalently}) \;\; 
\frac{dH}{d\phi} = \h \sqrt{H^2 - \frac{{\wti W}(\phi)}{H^2}} \; ,
\lab{vp-eq-W} \\
V(\phi) + 24\chi {\wti W}(\phi) = 4H^2 + 2 \frac{{\wti W}(\phi)}{H^2}
\lab{V-shifted-W} \; .
\er
The explicit functional dependence of $V(\phi)$ on $\phi$ is uniquely fixed by
Eq.\rf{V-shifted-W} in terms of the given ${\wti W}(\phi)$ (Fig.3 below) upon
substituting there the solution $H=H(\phi)$ of the second Eq.\rf{vp-eq-W}.

Let us now choose ${\wti W}(\phi)$ of the form qualitatively depicted on
Fig.3. 

\begin{figure}
\begin{center}
\includegraphics[width=10cm,keepaspectratio=true]{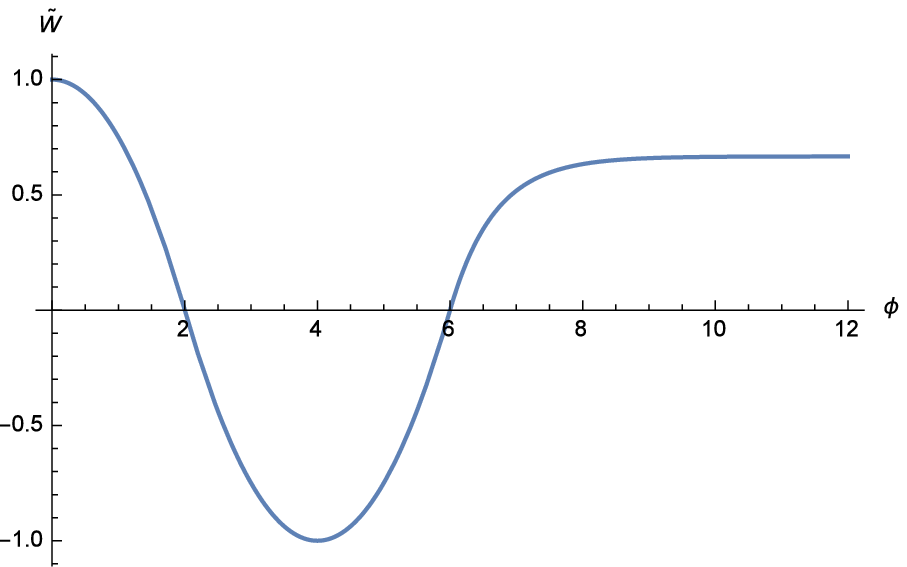}
\caption{Qualitative shape of ${\wti W}(\phi)$.}
\end{center}
\end{figure}

Since ${\wti W}(\phi)$ quickly approaches its asymptotic value
${\wti W}(\infty) >0$ for large $\phi$, Eqs.\rf{GB-FRLW-W}-\rf{vp-eq-W} will
have approximate solutions as in \rf{H-sol},\rf{vp-sol} where $M$ is replaced by
${\wti W}(\infty)$, \textsl{i.e.}, again we will have Big Bang at 
$t_{\rm BB} \simeq t_0 - \frac{\pi}{4\bigl({\wti W}(\infty)\bigr)^{1/4}}$.
On the other hand, for late times from \rf{GB-FRLW-W}-\rf{V-shifted-W} we
obtain:
\breq
H(t\to \infty) \simeq \bigl({\wti W}(0)\bigr)^{1/4} \quad ,\quad
\phi (t\to \infty) \simeq 0 \; ,
\lab{H-vp-late-W} \\
V(0) + 24\chi {\wti W}(0) = 6 \sqrt{{\wti W}(0)} 
= {\rm dark ~energy ~density} \; .
\lab{DE-W}
\er

In a subsequent study we will explore more systematically the more realistic
Einstein-Gauss-Bonnet model \rf{TMT-GB-W} which should involve numerical
treatment of the nonlinear system of equations \rf{GB-FRLW-W}-\rf{vp-eq-W}.

\section*{Acknowledgements}
We gratefully acknowledge support of our collaboration through 
the academic exchange agreement between the Ben-Gurion University in Beer-Sheva,
Israel, and the Bulgarian Academy of Sciences. 
E.N. and E.G. have received partial support from European COST actions
MP-1405 and CA-16104, and from CA-15117 and CA-16104, respectively.
E.N. and S.P. are also thankful to Bulgarian National Science Fund for
support via research grant DN-18/1. 
Finally, we are thankful to the referee for remarks contributing to improvements in
the text.


\end{document}